# Non-uniform wave momentum bandgap in biaxial anisotropic photonic time crystals

Junhua Dong,[*] Sihao Zhang,[*] Huan He, Huanan Li,[†] and Jingjun Xu[‡]

*MOE Key Laboratory of Weak-Light Nonlinear Photonics, School of Physics, Nankai University, Tianjin 300071, China*

*Photonic time crystals (PTCs) host momentum bandgaps enabling intriguing non-resonant light amplification in propagating waves, but opening substantial bandgaps demands refractive index changes too extreme for conventional nonlinear optics. Here, we introduce momentum bandgaps for non-uniform waves, including evanescent and ghost types, by extending PTCs to biaxial anisotropic photonic time crystals that periodically alternate between uniform biaxial anisotropy and isotropic media over time. We show that ghost waves, unlike evanescent waves, sustain only momentum bandgaps, opening wide bandgaps at even the smallest modulation depths. Moreover, we demonstrate momentum bandgap effects on non-uniform waves that can be amplified, or through decaying modes, selectively attenuated. We find that ghost wave momentum bandgaps uniquely boost refracted over reflected waves under one-way incidence, in stark contrast to balanced amplification seen in both propagating and evanescent waves. Our approach expands time-varying metamaterials by integrating wave characteristics, bridging the gap between conventional nonlinear optics and PTC momentum bandgaps, and shedding new light on extreme manipulation of surface polaritons.*

---

[*] These authors contributed equally to this work.
[†] hli01@nankai.edu.cn
[‡] jjxu@nankai.edu.cn

Photonic time crystals (PTCs) arise when material properties of a uniform medium are modulated periodically in time [1]-[5]. As plane waves propagate within PTCs, they experience time reflection and refraction while conserving momentum (wavevector $\vec{k}$) [6]-[7], leading to the interference of the reflected and refracted waves, that creates a band structure characterized by distinct momentum bands and bandgaps. These momentum bandgaps, inherent to PTCs, enable intriguing non-resonant light amplification by drawing energy from the modulation, bypassing phase-matching [8] and exhibiting possibilities for tunable lasing and transformed light-matter interactions [9]-[12]. Opening a noticeable momentum bandgap in PTCs requires large and ultrafast changes in the medium's refractive index [13]. Such substantial index variations are beyond the reach of most conventional nonlinear optical approaches, which, despite their instantaneous response, are orders of magnitude too weak [14]. Therefore, it is anticipated that, for now, the observation of momentum bandgaps [15]-[16] and their associated non-resonant amplification effects [17] remains limited to the microwave domain.

Momentum bandgaps in PTCs result from the dynamic interaction between waves and a time-modulated material platform. Yet, current research on PTC-capable materials has been concentrated on propagating waves, with little consideration given to the wave characteristics essential to PTC functionality. A concept distinct from propagating waves is that of non-uniform plane waves [18], such as the canonical example of evanescent waves and the newly identified ghost waves in transparent biaxial anisotropic dielectrics [19]-[21]. The former, with a purely imaginary wavevector component, is pivotal in applications ranging from efficient optical manipulation [22] to sub-diffraction imaging [23]-[24] and super-Planckian thermal emission [25]-[27]. The latter, possessing a complex-valued wavevector component with nonzero real and imaginary parts, merges the properties of propagating and evanescent waves and holds promise for applications in nonlinear optics and optical sensing [28]-[30]. Currently, the impact of time-modulated systems on non-uniform plane waves, especially the momentum bandgap phenomena for such waves within PTCs, remains elusive.

Here, we investigate the interactions of non-uniform plane waves with biaxial anisotropic photonic time crystals (B-APTCs), generated by switching in time between a lossless, non-magnetic, uniform biaxial anisotropic crystal and an isotropic one with a periodic (relative) permittivity tensor $\bar{\bar{\varepsilon}}(t) = \bar{\bar{\varepsilon}}(t + T)$, where $T$ is the modulation period. We find that while evanescent waves exhibit both momentum band and bandgaps like propagating waves, ghost waves uniquely manifest momentum bandgaps and notably, a broad bandgap can be opened with arbitrarily small modulation depths. The bandgaps can be utilized to amplify evanescent and ghost waves, and, through tailored interactions between incident forward and backward waves, to specifically attenuate them. Moreover, ghost waves, unlike evanescent and propagating waves, can leverage momentum bandgaps to break flux conservation constraint, leading to an enhancement of refracted over reflected waves under one-way incidence.

*Non-uniform waves in biaxial anisotropic crystals* — Let's begin our discussion by examining the generation of non-uniform waves in the B-APTC structures. During the initial time interval, we assume the propagation of a time-harmonic plane wave of the form $\propto \exp[i(\vec{k} \cdot \vec{r} - \omega_{inc}t)]$ through a biaxial crystal [31], described by the permittivity tensor $\bar{\bar{\varepsilon}}^{(1)} = \text{diag}(\varepsilon_x, \varepsilon_y, \varepsilon_z)$. The mode dispersion relation links the wave vector components $k_\alpha, \alpha = x, y, z$ to the angular frequency $\omega_{inc}$ and is given by [32]-[33]

$$k_z^2 = \frac{1}{2}\left[\left(\frac{\omega_{inc}}{c_0}\right)^2 (\varepsilon_x + \varepsilon_y) - \left(1 + \frac{\varepsilon_x}{\varepsilon_z}\right) k_x^2 - \left(1 + \frac{\varepsilon_y}{\varepsilon_z}\right) k_y^2 \pm \sqrt{Q}\right], \quad (1)$$

where the discriminant $Q$ reads

$$Q = \left[\left(\frac{\omega_{inc}}{c_0}\right)^2 (\varepsilon_x - \varepsilon_y) + \left(1 - \frac{\varepsilon_x}{\varepsilon_z}\right) k_x^2 - \left(1 - \frac{\varepsilon_y}{\varepsilon_z}\right) k_y^2\right]^2 + 4\left(1 - \frac{\varepsilon_x}{\varepsilon_z}\right)\left(1 - \frac{\varepsilon_y}{\varepsilon_z}\right) k_x^2 k_y^2,$$

and $c_0$ denotes the speed of light in free space. In Fig. **1**(a), we present the dispersion diagram from Eq. (1), using primary permittivity tensor components $\varepsilon_x = 2$, $\varepsilon_y = 4$, and $\varepsilon_z = 3$, and real (in-plane) wave vector components $k_x = 0.5k_0$ and $k_y = 0.32k_0$ [33], where $k_0 = \omega_0/c_0$ and $\omega_0 = 2\pi/T$. As the incident frequency $\omega_{inc}$ varies, three distinct wave types arise: propagating and evanescent waves when $k_z^2$ from Eq. (1) is positive or negative, respectively, when the discriminant $Q > 0$; and ghost waves when $k_z^2$ becomes complex with non-zero real and imaginary parts for $Q < 0$. Indeed, the critical frequencies $\omega_-$ and $\omega_+$, satisfying $k_z = 0$, along with $\omega_{c-}$ and $\omega_{c+}$, where $Q = 0$, define different frequency regimes highlighted by colored backgrounds in Fig. **1**(a). Apart from propagating wave regimes, evanescent waves occur between $\omega_-$ and $\omega_+$ or 0 and $\omega_{c-}$, and ghost waves between $\omega_{c-}$ and $\omega_{c+}$. Selecting incidence frequencies $\omega_{inc}/\omega_0$ as 0.336, 0.338, and 0.334, with corresponding $k_z/k_0$ values of 0.038, $i0.0215$, and $0.092 + i0.04$ [see the stars in the inset of Fig. **1**(a) on the solution branch of Eq. (1), where $\text{Re } k_z$, $\text{Im } k_z \geq 0$ [33], marked by solid (empty) symbols for real (imaginary) parts], we generate propagating, evanescent, and ghost waves, respectively, as illustrated in Fig. **1**(b). Evanescent and ghost waves, which have non-zero $\text{Im } k_z$, exhibit non-uniform behaviors: the former (with $\text{Re } k_z = 0$) decay exponentially, while the latter (with $\text{Re } k_z \neq 0$) oscillate as they decay, unlike propagating waves that have real wave vectors and sustain unattenuated oscillations.

***Band structure of B-APTCs for non-uniform waves*** —As depicted schematically in Fig. **2**(a), the B-APTC comprises a biaxial crystal with the permittivity tensor $\bar{\bar{\varepsilon}}^{(1)}$ and an isotropic medium with $\bar{\bar{\varepsilon}}^{(2)} = \varepsilon \bar{\bar{I}}_3$, where the superscript $(m)$, with $m = 1, 2$, labels the different media and $\bar{\bar{I}}_N$ is the $N$th order identity matrix. In the B-APTC initial temporal slab (biaxial crystal), we consider the specific $k_z$ solution branch of Eq. (1), as indicated in Fig. 1(a) via symbols, for the incident wave with frequency $\omega_{inc}$ and in-plane $k_x = 0.5k_0$, $k_y = 0.32k_0$ components. Due to the preserved spatial uniformity in the B-APTC, the generally complex wave vector $\vec{k} = [k_x, k_y, k_z]^T$ [where the superscript $T$ denotes transpose operation] is conserved. For the given $\vec{k}$ in Fig. 1(a), the dispersion relation in Eq. (1) for medium $m = 1$ (biaxial crystal) yields four distinct frequencies $\pm\omega_n^{(1)}$, $n = 1,2$, where $\omega_1^{(1)}$ corresponds to the incident frequency $\omega_{inc}$. These four frequencies result in two pairs of plane waves (for $n = 1, 2$), each consisting of a forward and a backward wave, sharing an electric displacement vector $\vec{d}_n$ given by [36]

$$\vec{d}_n = \left[\frac{\varepsilon_x k_x k c_0^2}{k^2 c_0^2 - \left(\omega_n^{(1)}\right)^2 \varepsilon_x}, \frac{\varepsilon_y k_y k c_0^2}{k^2 c_0^2 - \left(\omega_n^{(1)}\right)^2 \varepsilon_y}, \frac{\varepsilon_z k_z k c_0^2}{k^2 c_0^2 - \left(\omega_n^{(1)}\right)^2 \varepsilon_z}\right]^T, \qquad (2)$$

where the complex wave number $k$ is defined as $k = \sqrt{k_x^2 + k_y^2 + k_z^2}$ [37]. In the isotropic medium ($m = 2$) of the B-APTC, the vectors $\vec{d}_n$ for $n = 1, 2$ from Eq. (2) also provide a basis for both forward and backward waves with degenerate frequencies $\omega_1^{(2)} = \omega_2^{(2)} = kc_0/\sqrt{\varepsilon}$ and their negative counterparts. Consequently, the time-dependent total displacement field $\vec{D}(\vec{r}, t)$ in the B-APTC for the given wave vector $\vec{k}$ can be expressed as

$$\vec{D}(\vec{r}, t)/\varepsilon_0 = e^{i\vec{k}\cdot\vec{r}} \sum_{n=1}^{2} \vec{d}_n [1, \quad 1] \psi_n(t) + c.c., \qquad (3)$$

where $\varepsilon_0$ is the vacuum permittivity, $c.c.$ stands for complex conjugate, and the column vectors $\psi_n(t) = [f_n(t), \quad b_n(t)]^T$ contain the time-dependent amplitudes $f_n(t) \propto e^{-i\omega_n^{(m)} t}$ and $b_n(t) \propto e^{i\omega_n^{(m)} t}$ associated with the frequencies $\pm\omega_n^{(m)}$ for the forward and backward waves, respectively, in medium $m$ [38]. Depending on the sign of $\pm\omega_n^{(m)}$, the phase velocities of the forward and backward waves for a common $\vec{k}$ are opposite, aligning with or against $\operatorname{Re}\vec{k}$, yet both share the same attenuation direction, $\operatorname{Im}\vec{k} \parallel \vec{z}$, normal to the amplitude planes [33].

The state vector $\psi(t) \equiv \begin{bmatrix} \psi_1(t) \\ \psi_2(t) \end{bmatrix}$, determining the total displacement field $\vec{D}(\vec{r},t)$ in Eq. (3), evolves within the B-APTC according to the temporal transfer matrix method [11]. Specifically, for a temporal slab of medium $m = 1, 2$, each wave propagates freely and thus the evolution of $\psi(t)$ is governed by a diagonal matrix $F^{(m)} = \operatorname{diag}\{F_1^{(m)}, \ F_2^{(m)}\}$, where $F_n^{(m)} = \operatorname{diag}\{e^{-i\omega_n^{(m)}T^{(m)}}, \ e^{i\omega_n^{(m)}T^{(m)}}\}, n = 1, 2$ with $T^{(m)}$ being the duration of the slab. Upon encountering a time interface at $t_s$, where the medium transitions from $m = 1$ to $2$, waves with polarization $\vec{d}_1$ evolve independently from those with polarization $\vec{d}_2$, and a forward incident non-uniform wave oscillating at $\omega_1^{(1)} = \omega_{inc}$ scatters into forward and backward waves with frequencies $\pm\omega_1^{(2)}$ in medium $m = 2$, as schematically illustrated in Fig. 2(a) at the initial time interface $t_s = T^{(1)}$. Indeed, the continuity of the displacement field $\vec{D}$ and its time derivative $\partial\vec{D}/\partial t$ [39] requires that the abrupt change of $\psi(t)$ at $t_s$, $\psi(t_s^+) = J^{(2,1)}\psi(t_s^-)$, is described by the block-diagonal matching matrix $J^{(2,1)} = \operatorname{diag}\{J_1^{(2,1)}, J_2^{(2,1)}\}$ [33], with blocks $J_n^{(2,1)} \equiv \frac{1}{2\omega_n^{(2)}}\begin{bmatrix} \omega_n^{(2)} + \omega_n^{(1)}, & \omega_n^{(2)} - \omega_n^{(1)} \\ \omega_n^{(2)} - \omega_n^{(1)}, & \omega_n^{(2)} + \omega_n^{(1)} \end{bmatrix}$. Likewise, at the reverse time interface where the medium switches from $m = 2$ back to $1$, the matching matrix becomes $J^{(1,2)} = [J^{(2,1)}]^{-1}$. The combination of the block-diagonal propagation and matching matrices dictate the *independent* evolution of the state vector components $\psi_n(t), n = 1, 2$ in the B-APTC.

We focus on the evolution of $\psi_1(t)$, corresponding to the incident waves oscillating at frequency $\omega_{inc} = \omega_1^{(1)}$. The temporal transfer matrix $M_1(\vec{k})$ for a B-APTC unit cell, indicated in Fig. **2**(a) by the red dashed box, is $M_1(\vec{k}) = J_1^{(1,2)} F_1^{(2)} J_1^{(2,1)} F_1^{(1)}$. Its eigenvalues $\lambda_\pm$, satisfying $M_1(\vec{k})|\pm\rangle = \lambda_\pm |\pm\rangle$, can be expressed as $\lambda_\pm = e^{-i\Omega_\pm T}$ with $\Omega_\pm$ being the Floquet frequencies. By solving the secular equation $\det\{M_1(\vec{k}) - e^{-i\Omega_\pm T} \bar{I}_2\} = 0$, we obtain the B-APTC band structure, namely $\Omega_\pm$ versus $\vec{k}$. In Fig. **2**(b), we depict the band structure for incident waves in the biaxial crystal $\bar{\bar{\varepsilon}}^{(1)}$, related to Fig. **1**(a). Here, we set $T^{(1)} = T^{(2)} = 0.5T$ and $\varepsilon = 1$ for isotropic $\bar{\bar{\varepsilon}}^{(2)}$, and plot the complex Floquet frequencies $\Omega_\pm$ as a function of $\omega_{inc}$. As seen, the variation in $\omega_{inc}$ reveals both momentum band where $\text{Im}(\Omega_\pm) = 0$ (indicating time-oscillating Floquet modes without systematic growth or decay) and bandgap where $\text{Im}(\Omega_\pm) \neq 0$ (corresponding to Floquet modes that grow or decay exponentially over time). Similar to propagating waves, the B-APTC features momentum bands and bandgaps for evanescent waves, yet interestingly, it opens a distinct momentum bandgap in the case of ghost waves.

To understand the B-APTC band structure, we specify the unit-cell transfer matrix $M_1(\vec{k})$ as $M_1(\vec{k}) = \begin{bmatrix} a_+ & d_- \\ d_+ & a_- \end{bmatrix}$. Since $M_1(\vec{k})$ has a determinant of 1, we can obtain its eigenvalues, $\lambda_\pm$, as $\lambda_\pm = [(a_+ + a_-) \pm \sqrt{(a_+ + a_-)^2 - 4}]/2$. For the incidence of evanescent waves, the matrix elements of $M_1(\vec{k})$ exhibit conjugate symmetry—specifically, $a_+^* = a_-$ and $d_+^* = d_-$—akin to that of propagating waves. This symmetry arises because in such cases the frequency $\omega_1^{(2)}$ of the wave in the isotropic medium $m = 2$ is either real or purely imaginary, see [33]. Hence, the eigenvalues $\lambda_\pm$ reduce to $\lambda_\pm = \text{Re}\, a_+ \pm \sqrt{(\text{Re}\, a_+)^2 - 1}$, forming a momentum band when $(\text{Re}\, a_+)^2 < 1$ and thus $|\lambda_\pm| = 1$, and a bandgap when $(\text{Re}\, a_+)^2 > 1$ with $|\lambda_+| = 1/|\lambda_-| \neq 1$.

For ghost wave incidence, quite differently, the frequency $\omega_1^{(2)}$ becomes complex, resulting in $a_+^* \neq a_-$ and a generally non-zero imaginary part for $(a_+ + a_-)$. This discrepancy gives rise to a full momentum bandgap as the incident frequency $\omega_{inc}$ varies for ghost waves. Hence, the bandgap width in $\omega_{inc}$ is solely determined by the permittivity $\bar{\bar{\varepsilon}}^{(1)}$ of the biaxial crystal. By significantly reducing the modulation depth, $\Delta n = \max\limits_{\alpha=x,y,z}\{|\sqrt{\varepsilon_\alpha} - \sqrt{\varepsilon}|\}$, we can maintain the wide bandgap shown in Fig. **2**(b), albeit at the expense of reduced maximum amplification strength $g_{max} = \max\limits_{\omega_{inc}}\{|\text{Im}(\Omega_\pm T)|\}$, in contrast to the closed bandgaps for both evanescent and propagating waves, see Fig. **2**(c) and [33]. Unlike the method described in Ref. [40], this approach to widening the momentum bandgap with ghost waves does not necessitate the use of resonant materials. The transparent dielectric media in B-APTCs enables the observation of the diminished amplification effect for the ghost-wave momentum bandgaps.

*Momentum-bandgap-based manipulation of non-uniform waves in B-APTCs*—Momentum bandgaps can induce wave amplification through broken time-translation symmetry, contrasting with wave attenuation in energy (frequency) gaps of conventional photonic crystals. In the upper panels of Fig. **3**, we employ the B-APTC to showcase how wave amplification effects, associated with momentum bandgaps, can also encompass non-uniform waves. Specifically, we consider forward wave incidence by setting the initial state as $\psi_1(0^+) = [1, 0]^T$. The incident frequency $\omega_{inc}$ is chosen as $0.338\omega_0$ for evanescent waves [Fig. **3**(a)] and $0.334\omega_0$ for ghost waves [Fig. **3**(b)]. Therefore, both wave types fall within the B-APTC momentum bandgap [Fig. **2**(b)], with wave amplification evident in the upper panels of Figs. **3**(a) and **3**(b), where theoretical predictions (solid lines) of the displacement field components $D_x(\vec{r} = 0, t)$ match numerical simulation (symbols). In these cases, both forward and backward waves grow over time, see the dashed lines for the $x$-component of their respective displacement fields $\vec{D}^\alpha = \vec{d}_1 e^{i\vec{k}\cdot\vec{r}}\alpha_1(t)\varepsilon_0 + c.c., \alpha = f, b$ [33]. A recent experiment with metasurface-based PTCs has indeed amplified surface-wave signals [17]; however, in these setups, the surrounding medium where the waves decay as evanescent waves is not time-modulated as it is in our B-APTCs.

To elucidate these characteristics, we employ the complete basis formed by the eigenvectors $|\pm\rangle$ of the unit-cell transfer matrix $M_1(\vec{k})$ to expand the initial state $\psi_1(0^+)$ as $\psi_1(0^+) = \sum_{\sigma=\pm} p_\sigma |\sigma\rangle$, where $p_\sigma$ are the expansion coefficients. After $N$ modulation cycles, the final state at time $NT^+$ becomes $\psi_1(NT^+) = \sum_{\sigma=\pm} p_\sigma e^{-iN\Omega_\pm T}|\sigma\rangle$ by applying the $N$th power of $M_1(\vec{k})$ to the initial state $\psi_1(0^+)$. In a momentum bandgap, both growing and decaying Floquet modes corresponding to the eigenvectors $|\pm\rangle$ are present, each exhibiting complex Floquet frequencies satisfying $\Omega_+ = -\Omega_-$ [see Fig. 2(b)]. As $N$ increases, the exponentially growing Floquet mode becomes dominant, overshadowing the decaying mode and amplifying waves propagating in both directions. By specifically aligning the initial state $\psi_1(0^+)$ with the decaying mode, we can also induce the decay of non-uniform waves within the momentum bandgaps. As shown in the lower panels of Fig. **3**, when $\psi_1(0^+)$ takes $[1, 0.5013 + i0.8652]^T$, both forward and backward waves (dashed lines), as well as the total field (solid lines), exhibit exponential decay for evanescence wave incidence [Fig. **3**(a)], and similarly, when $\psi_1(0^+) = [1, 0.5288 + i0.8958]^T$ for ghost wave incidence [Fig. **3**(b)]. This wave suppression method leverages initial counterpropagating waves, extending the approach of photonic collisions at time interfaces [41]-[42] to encompass the dynamics of PTC momentum bandgaps and non-uniform wave regimes.

For propagating waves in PTCs, a general constraint exists that ensures the conservation of Poynting flux across any sequence of modulation cycles [43]. This results in a precise balance between forward and backward wave components, a phenomenon that persists even within momentum bandgaps. Ghost waves, as a unique type of non-uniform wave, present an exception where this conservation constraint can be broken. To see this, we examine the (time-averaged) Poynting vectors $\vec{S}$ for non-uniform waves in the biaxial crystal $\bar{\bar{\varepsilon}}^{(1)}$ of the B-APTC. The Poynting vector $\vec{S}$, associated with the state vector $\psi_1(t)$ with amplitude $\alpha_1(t)$ for both forward ($\alpha = f$) and backward ($\alpha = b$) waves, is determined by the net energy flux $\vec{S} = \vec{S}^f - \vec{S}^b$ with

$$\vec{S}^\alpha = |\alpha_1(t)|^2 \frac{2\operatorname{Re}(\vec{e}_1 \times \vec{h}_1^*)}{\eta_0} e^{-2\operatorname{Im}(k_z)z}, \tag{4}$$

where $\vec{e}_1 = [\bar{\bar{\varepsilon}}^{(1)}]^{-1}\vec{d}_1$ and $\vec{h}_1 = (\vec{k}c_0 \times \vec{e}_1)/\omega_1^{(1)}$ are the electric and magnetic field basis vectors linked to the displacement vectors $\vec{d}_1$ in Eq. (2), $\eta_0$ is the impedance of free space. Evanescent and ghost waves carry no energy in their out-of-plane decay direction, rendering the $z$ component of the Poynting vector $\vec{S}$ as zero (i.e., $S_z = 0$), due to the vanishing cross-product component $\operatorname{Re}(\vec{e}_1 \times \vec{h}_1^*)_z$ [see Eq. (4)]. However, the Poynting vector component $S_\parallel$ along the $x - y$ plane remains nonzero, indicating the presence of an in-plane energy flux.

Assuming an initial forward wave incidence at time $t = 0^+$, the normalized in-plane energy flux after $N$ modulation cycles, $\hat{S}_\parallel(NT^+) \equiv S_\parallel(NT^+)/|S_\parallel(0^+)|$, is represented by the difference $\hat{S}_\parallel^f(NT^+) - \hat{S}_\parallel^b(NT^+)$, where $\hat{S}_\parallel^\alpha(NT^+) = |\alpha_1(NT^+)|^2/|f_1(0^+)|^2$, accounting for the relative contributions of forward ($\alpha = f$) and backward ($\alpha = b$) wave components. For evanescent waves, the normalized energy flux $\hat{S}_\parallel(NT^+) = 1$ irrespective of the number of modulation cycles $N$, similar to propagating waves, reflecting the shared properties of the transfer matrix $M_1^N$ with $\det M_1^N = 1$ and the conjugate symmetry in its elements like $M_1$ (see also [33]). In contrast, for ghost waves, the absence of conjugate symmetry in the transfer matrix elements signifies a departure from flux conservation, extending the momentum-bandgap-based wave manipulation beyond conventional constraints. As shown in Fig. **4**(a) with $N = 10$ modulation cycles, $\hat{S}_\parallel(NT^+)$ can largely exceed 1 as the incident frequency $\omega_{inc}$ varies in the momentum bandgap for ghost waves. Furthermore, the sign of $\hat{S}_\parallel(NT^+)$ for large $N$ does not change when switching to initial backward wave incidence (see [33]), indicating directional amplification reminiscent of unidirectional propagation in nonreciprocal structures [44]-[45].

For $\omega_{inc}/\omega_0 = 0.150, 0.320,$ and $0.334$, Fig. **4**(b) illustrates the exponential increase of $\hat{S}_\parallel(NT^+)$ over $N$ at variable rates, with the insets highlighting the imbalance between the forward and backward wave components, $\hat{S}_\parallel^f$ and $\hat{S}_\parallel^b$, for each case, as they are amplified in the momentum bandgap. The relative difference between the amplified forward and backward wave components becomes considerable at $\omega_{inc}/\omega_0 = 0.150$ [Fig. **4**(b): left inset], and this disparity in amplification extends over a broad incident frequency range within the bandgap, as indicated by the dashed blue line in Fig. **4**(a) using the measure $R_\parallel(NT^+) = S_\parallel(NT^+)/|f_1(NT^+)|^2$. This unbalanced amplification, driven by ghost wave momentum bandgaps, can extend to surface mode manipulation [33]. The resulting unbounded growth of surface waves, assuming no nonlinearity, indicates instability [46], contrasting with the wave control methods employed in time-varying systems operating in stable regimes, as exemplified in Refs. [47]-[48].

*Conclusions* — In this paper, we have expanded PTCs into the domain of B-APTCs and examined their interaction with non-uniform plane waves, focusing on the momentum-bandgap phenomena for evanescent and ghost waves. We found that evanescent waves exhibit momentum bandgaps, while ghost waves present them uniquely and can form broad bandgaps at minimal modulation depths. These bandgaps can amplify, or via the decaying mode, attenuate both evanescent and ghost waves. Furthermore, ghost waves—unlike evanescent waves—can exploit momentum bandgaps to boost refracted over reflected waves under one-way incidence, diverging from the balanced amplification in propagating waves within PTC bandgaps due to flux conservation. Our approach bridges the gap between traditional nonlinear optics and PTC momentum bandgaps, enriching four-dimensional optics [49] with time-varying metamaterials that integrate wave characteristics, potentially extending to surface polariton manipulation.

*Acknowledgements* 一 This work was supported by the National Natural Science Foundation of China (Grant Nos. 12274240, 92250302).

**Figures**

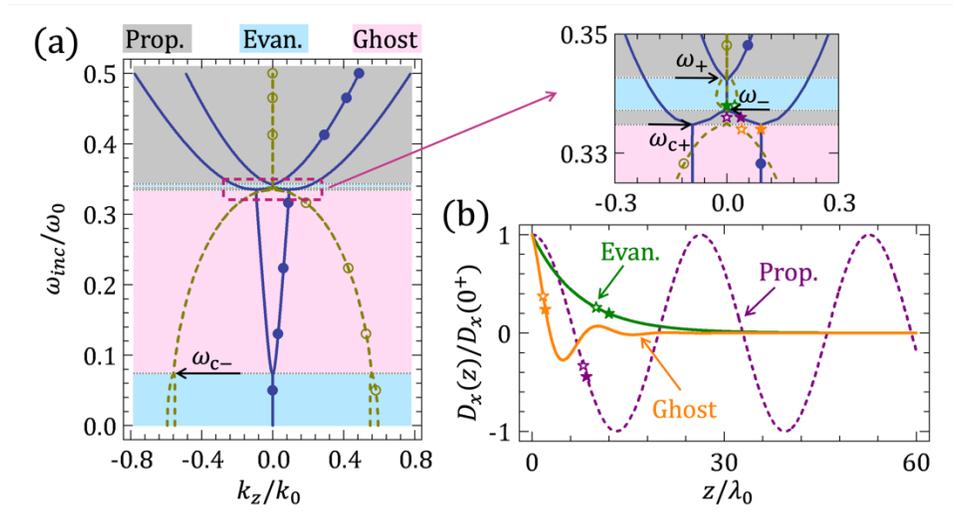

**Fig. 1.** (a) Dispersion diagram for biaxial crystal $\bar{\bar{\varepsilon}}^{(1)}$ with $\varepsilon_x = 2$, $\varepsilon_y = 4$, and $\varepsilon_z = 3$, showing plane waves at incident frequency $\omega_{inc}$ and complex wave vector component $k_z$ for fixed $k_x = 0.5k_0$ and $k_y = 0.32k_0$. Solid and dashed lines show the real and imaginary parts of $k_z$, respectively, with symbols marking the branch for incident waves in the B-APTC. Critical frequencies $\omega_{c\pm}$ and $\omega_{\pm}$ delineate the regimes for propagating, evanescent, and ghost waves, as shown by distinct colored backgrounds. (b) The displacement field component $D_x$ for the three

different wave types in the biaxial crystal, corresponding to the distinct stars in the inset of (a), as a function of position $z$ (in units of $\lambda_0 = 2\pi/k_0$).

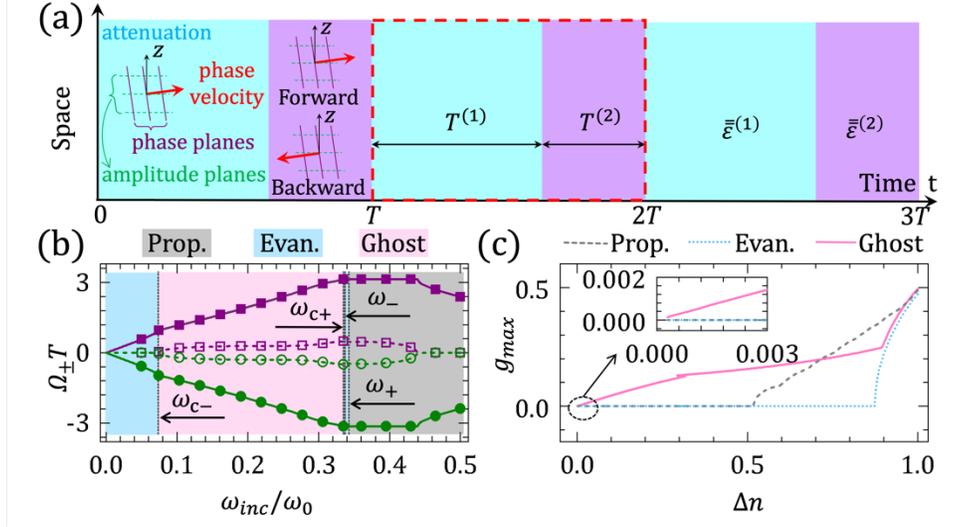

**Fig. 2.** (a) Schematic of the B-APTC and the scattering of a forward incident non-uniform wave at the initial time interface. (b) Band structure of the B-APTC, showing normalized Floquet frequencies $\Omega_\pm T$ plotted against the incident frequency $\omega_{inc}$ in the initial temporal slab of the biaxial crystal $\bar{\bar{\varepsilon}}^{(1)}$ [see Fig. **1**(a)]. Real and imaginary parts are given by lines of the same color, using solid symbols for the real part and corresponding empty ones for the imaginary part. Colored backgrounds denote various wave regimes, correlating with Fig. **1**(a). Parameters settings include $T^{(1)} = T^{(2)} = 0.5T$, and $\varepsilon = 1$ for $\bar{\bar{\varepsilon}}^{(2)}$, with other parameters matching those in Fig. **1**(a). (c) Maximum amplification strength $g_{max}$ of momentum bandgaps for propagating, evanescent, and ghost waves, as shown in (b), versus the modulation depth $\Delta n$.

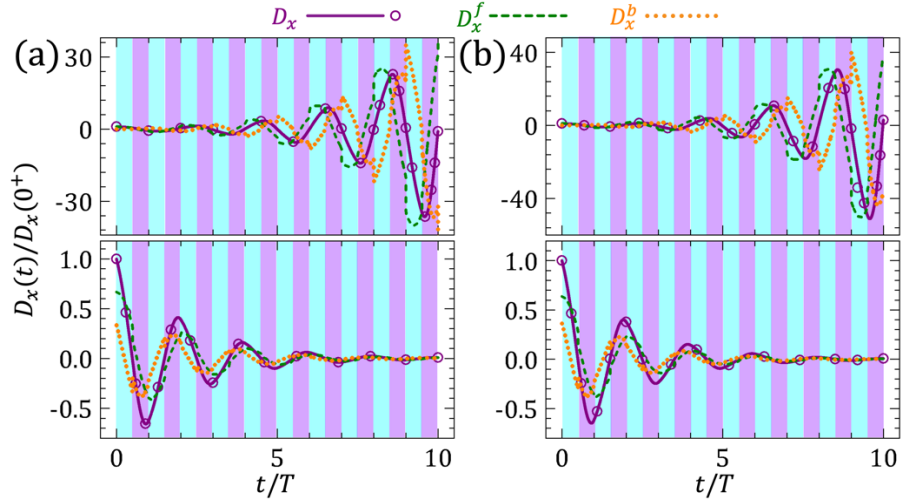

**Fig. 3.** Time evolution of the normalized total $(D_x)$, forward $(D_x^f)$, and backward $(D_x^b)$ electric displacement components at the origin, for (a) evanescent wave incident at $\omega_{inc} = 0.338\omega_0$, and (b) ghost wave incidence at $\omega_{inc} = 0.334\omega_0$, in the B-APTC momentum bandgap [see Fig. 2(b)]. The upper panels illustrate initial states $\psi_1(0^+) = [1, 0]^T$ for both (a) and (b), while the lower panels depict initial states $\psi_1(0^+) \approx [1, 0.5013 + i0.8652]^T$ for (a) and $[1, 0.5288 + i0.8958]^T$ for (b). Solid lines show theoretical predictions of $D_x$, with symbols indicating numerical results from time-domain Maxwell's equations. Colored backgrounds and parameters employed are in accordance with those presented in Fig. 2(a, b).

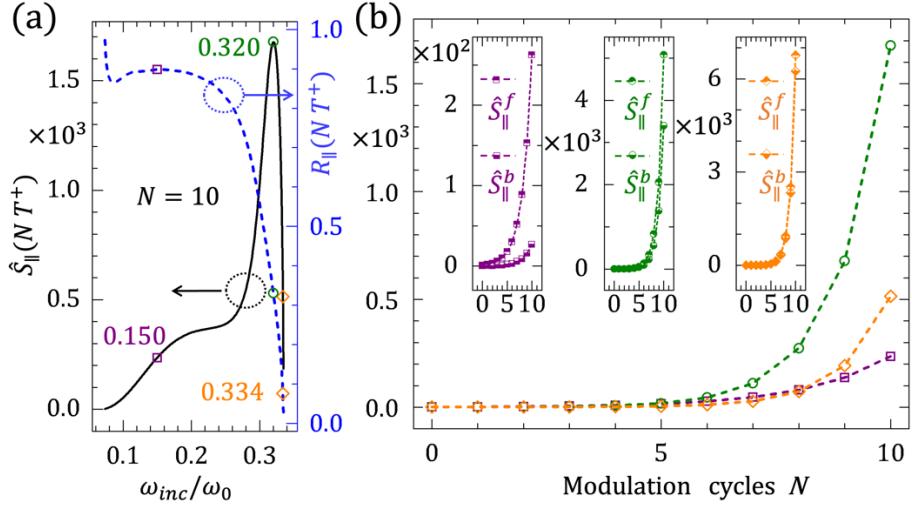

**Fig. 4.** (a) At $N = 10$ modulation cycles, the normalized in-plane energy flux $\hat{S}_\parallel(NT^+)$ (solid black line) and the relative forward-backward wave contribution $R_\parallel(NT^+)$ (dashed blue line) as a function of $\omega_{inc}$ within the ghost wave momentum bandgap [see Fig. 2(b)], with initial forward wave incidence at time $0^+$. (b) $\hat{S}_\parallel(NT^+)$ versus $N$ for ghost wave incidence at frequencies $\omega_{inc}/\omega_0 = 0.150$ (squares), $0.320$ (circles), and $0.334$ (diamonds) [see also (a)], with the insets for the contribution of forward ($\hat{S}_\parallel^f$) and backward ($\hat{S}_\parallel^b$) wave components.